\documentclass[sigconf]{acmart}

\AtBeginDocument{%
  }

\copyrightyear{2026}
\acmYear{2026}
\setcopyright{cc}
\setcctype{by}
\acmConference[CHI EA '26]{Extended Abstracts of the 2026 CHI Conference on Human Factors in Computing Systems}{April 13--17, 2026}{Barcelona, Spain}
\acmBooktitle{Extended Abstracts of the 2026 CHI Conference on Human Factors in Computing Systems (CHI EA '26), April 13--17, 2026, Barcelona, Spain}
\acmDOI{10.1145/3772363.3799162}
\acmISBN{979-8-4007-2281-3/2026/04}

\usepackage{subcaption}

\begin{document}

% ----------------------------------------------------------------------
% Title + Authors
% ----------------------------------------------------------------------
\title{The Pen: Episodic Cognitive Assistance via an Ear-Worn Interface}

\author{Yonatan Tussa}
\affiliation{%
  \institution{University of Maryland, College Park}
  \city{College Park}
  \state{MD}
  \country{USA}
}
\email{ytussa@umd.edu}

\author{Andy Heredia}
\affiliation{%
  \institution{University of Maryland Global Campus}
  \city{Adelphi}
  \state{MD}
  \country{USA}
}
\email{aheredia7@umgc.edu}

% Recommended when author list is short: still define to avoid header overflow issues later
\renewcommand{\shortauthors}{Tussa and Heredia}

% ----------------------------------------------------------------------
% Abstract, CCS, Keywords
% ----------------------------------------------------------------------
\begin{abstract}
Wearable AI is often designed as always-available, yet continuous availability can conflict with how people work and socialize, creating discomfort around privacy, disruption, and unclear system boundaries. This paper explores episodic use of wearable AI, where assistance is intentionally invoked for short periods of focused activity and set aside when no longer needed, with a form factor that reflects this paradigm of wearing and taking off a device between sessions. We present The Pen, an ear-worn device resembling a pen, for episodic, situated cognitive assistance. The device supports short, on-demand assistance sessions using voice and visual context, with clear start/end boundaries and local processing. We report findings from an exploratory study showing how layered activation boundaries shape users’ sense of agency, cognitive flow, and social comfort.
\end{abstract}




% ----------------------------------------------------------------------
% Teaser
% ----------------------------------------------------------------------
\begin{teaserfigure}
  \includegraphics[width=\textwidth]{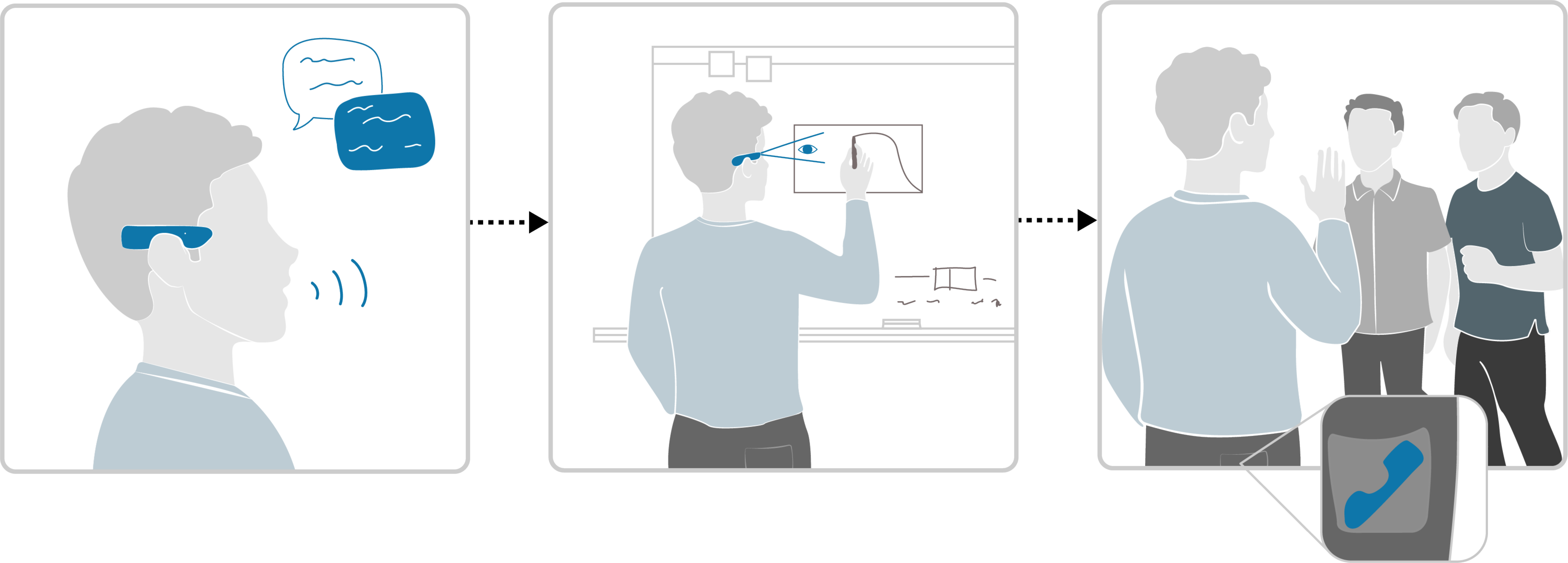}
  \caption{The Pen is worn behind the ear and the user is speaking to it (left); user continues whiteboarding while the device observes visual context (middle); user disengages by removing the device and placing it in a pocket before resuming social interaction (right).}
  \Description{A teaser figure illustrating The Pen as an ear-worn interface. The user speaks to the device while wearing it. The user then whiteboards while the device remains behind the ear, supporting an in-the-moment assistance episode. Finally, the user removes the device when approached by others, signaling the end of the assistance session.}
  \label{fig:teaser}
\end{teaserfigure}

\maketitle

% ----------------------------------------------------------------------
% Body
% ----------------------------------------------------------------------
\section{Introduction}

Wearable computing promises in-the-moment assistance during everyday tasks, yet most wearable AI systems assume continuous availability---always listening, sensing, and ready for interaction. In practice, persistent sensing raises privacy concerns for wearers and bystanders, introduces ambiguity around when data is being collected, and can undermine users’ sense of agency and control over the system’s operation \cite{orchard2022smartglasses, iqbal2023smartglasses, bhardwaj2024infocusoutofprivacy, alharbi2018cantbemyself, denning2014bystanders, bellotti2001intelligibility, bajorunaite2024vrheadsets}.

A recurring limitation of prior wearable AI systems is their assumption that intelligence should operate continuously in the background. These designs blur important boundaries: when the system is active, what it accesses, and how its involvement is legible to others. As a result, systems can feel intrusive. Existing interaction paradigms offer limited relief, with smartphones and laptops requiring shifts in gaze and attention that can disrupt cognitive flow. Voice assistants reduce visual demand but can be socially exposed, especially in shared workspaces. At the same time, ear-worn platforms are becoming increasingly capable sensing and interaction devices, enabling inference from audio and motion signals in everyday form factors \cite{hu2024iatrack, roddiger2022earablesurvey}.

Rather than defaulting to ambient, always-on intelligence, we explore episodic wearable assistance where users wear a device for short, task-bounded sessions and disengage when the task ends. This approach takes inspiration from foundational HCI work on intelligibility and accountability in context-aware systems \cite{bellotti2001intelligibility}, and on research showing that the act of wearing itself shapes expectations, meaning, and social interpretation \cite{orchard2022smartglasses, koelle2019socialacceptability}. This aligns with a growing body of work on proactive and assistive agents that must carefully manage when and how they intervene \cite{compeer2024, lee2025sensibleagent}.

We investigate this paradigm with The Pen, an ear-worn wearable designed for short, task-bounded assistance sessions. Users put the device behind their ear at the start of a task and remove it when the task ends, using the act of wearing to clearly mark the beginning and end of an assistance episode.

Our contribution centers on interaction design: we formalize episodic wearable assistance as a multi-layered boundary problem spanning physical, interactional, and perceptual cues.

\paragraph{We ask the following questions:}
\begin{itemize}
    \item How are episodic boundaries formed in practice?
    \item How do these boundaries influence users’ sense of agency and cognitive flow?
    \item What cues help users understand what the system is doing?
\end{itemize}

\paragraph{Our contributions are the following:}
\begin{itemize}
  \item \textbf{Concept:} We introduce episodic wearable assistance as an alternative to always-on wearable AI and a multi-layered boundary problem.
  \item \textbf{System:} We present The Pen, an ear-worn pen-shaped device designed to provide episodic assistance.
  \item \textbf{Study:} We report findings on episodic wearable interaction and tensions between agency, legibility, and social comfort.
\end{itemize}

\paragraph{Relationship to Prior Work.}
This paper builds on our prior work \cite{tussa2025lessonslearneddevelopingprivacypreserving}, which examined the systems and hardware integration challenges involved in building a multimodal wearable capable of local voice-and-vision inference. Here we focus on interaction design and user experience of episodic, task-bounded assistance.

\paragraph{Episodic boundaries as multi-layered interaction.}
In this paper, we use the term "episodic" to mean assistance that is intentionally entered and exited for a bounded period, rather than continuously available in the background. Our findings suggest episodic interaction boundaries are not solely tied to physical boundaries, but are co-produced through a physical ritual, an explicit activation action, and perceptible feedback that makes state legible. We refer to these as (1) physical boundaries (wear/remove), (2) interaction boundaries (a deliberate trigger such as pressing), and (3) perceptual boundaries (audio/haptic cues that communicate system state to the user).

% \paragraph{Deixis.}
% We draw on the concept of deixis to interpret how wearable systems communicate state through embodied action. Deictic signals are contextual or indexical cues whose meaning depends on the present interaction (e.g., “here,” “now,” or a pointing gesture). Prior work in ubiquitous computing has explored hands-free deixis through gaze-based interaction, where eye movement functions as a contextual index for reference and control \cite{smith2005viewpointer}. In wearable systems, physical acts such as wearing, removing, or pressing a device can similarly function as deictic cues that situate the system within a specific moment of use.

\section{Related Work}

Our work draws on research in ear-worn and screenless wearables, situated cognitive assistance, and the social and ethical implications of wearable sensing. Across these areas, many systems assume persistent availability and continuous inference. In contrast, we focus on episodic use: assistance invoked for short, task-bounded moments and then disengaged, with interaction boundaries made legible through form factor and feedback.

\subsection{Ear-Worn and Screenless Wearable Interaction}

Prior work explores discreet, hands-free interaction through head and ear-worn devices. AlterEgo demonstrates silent-speech interaction as a pathway toward always-available computing \cite{kapur2018alterego}. Surveys and taxonomies of earables highlight the ear as a socially acceptable site for audio-first interaction while surfacing challenges around comfort, ergonomics, sensing reliability, and legibility \cite{roddiger2022earablesurvey, earableSurvey2025}. Recent work shows that commodity earphones can support inference beyond audio I/O (e.g., head motion tracking via fused IMU and acoustics) \cite{hu2024iatrack}. EarRumble further show how subtle bodily actions can enable hands and eyes-free control \cite{roddiger2021earrumble}.

Beyond earables, screenless wearables such as FingerTrak and IRIS demonstrate lightweight capture of task-relevant context without the user shifting attention to a phone \cite{fang2020fingertrak, kim2024iris}. FingerTrak shows how a wearable can infer task-relevant hand activity under partial visibility using sensing to recover interaction context and IRIS demonstrates a ring with a camera that captures visual context and offloads recognition to a paired device.

\subsection{Situated Assistance and Memory Support}

A growing body of work explores AI assistance embedded within ongoing activities. Systems such as Memoro use large language models to provide concise, real-time memory augmentation during work \cite{zulfikar2024memoro}, while other agents intervene during procedures or everyday tasks to offer guidance or corrective feedback \cite{arakawa2024prism, lee2025sensibleagent}. Related work on proactive conversational agents examines when and how systems should initiate support to remain helpful without becoming intrusive \cite{compeer2024}.

\subsection{Activation Tax and Micro-Interaction Costs}

Even when assistance is useful, the cost of invoking it can determine whether it fits into real work. Prior work quantifies the ``activation tax'' of mobile interaction, showing that simply initiating phone use can impose meaningful time and attention overhead, making micro-interactions sensitive to friction \cite{tan2010interfacesonthego}. This motivates our emphasis on episodic assistance with a distinct entry/exit: if users seek short bursts of help (e.g., clarification during reading), the interface must enable rapid engagement and disengagement without requiring posture shifts or repeated attempts.

\subsection{Privacy, Social Acceptability, and Legible Boundaries}

Wearable sensing raises longstanding concerns around bystander privacy, ambiguity, and the social implications of capture and inference \cite{denning2014bystanders, alharbi2018cantbemyself, orchard2022smartglasses}. Studies of camera-equipped wearables show that unclear system state and ambiguous recording boundaries can undermine trust and social comfort for both wearers and bystanders \cite{bhardwaj2024infocusoutofprivacy, koelle2019socialacceptability}. Cardea further argues that camera privacy is context-dependent and emphasizes mechanisms that allow users to manage capture and sharing in ways that match situational expectations \cite{shu2016cardeacontextawarevisualprivacy}. Technical approaches such as ScreenAvoider demonstrate methods for limiting unintended capture in ubiquitous sensing systems \cite{korayem2014screenavoider}.

Mirai describes an always-on wearable that continuously senses and delivers proactive assistance, illustrating both the promise of persistent inference and the risks of diminished legibility when systems are continuously listening and observing \cite{fang2025miraiwearableproactiveai}.

\section{The Pen}
\label{sec:thepen}

\begin{figure}[t]
  \centering
  \begin{subfigure}{0.48\linewidth}
    \centering
    \includegraphics[
      width=\linewidth,
      trim=0mm 40mm 0mm 75mm,
      clip
    ]{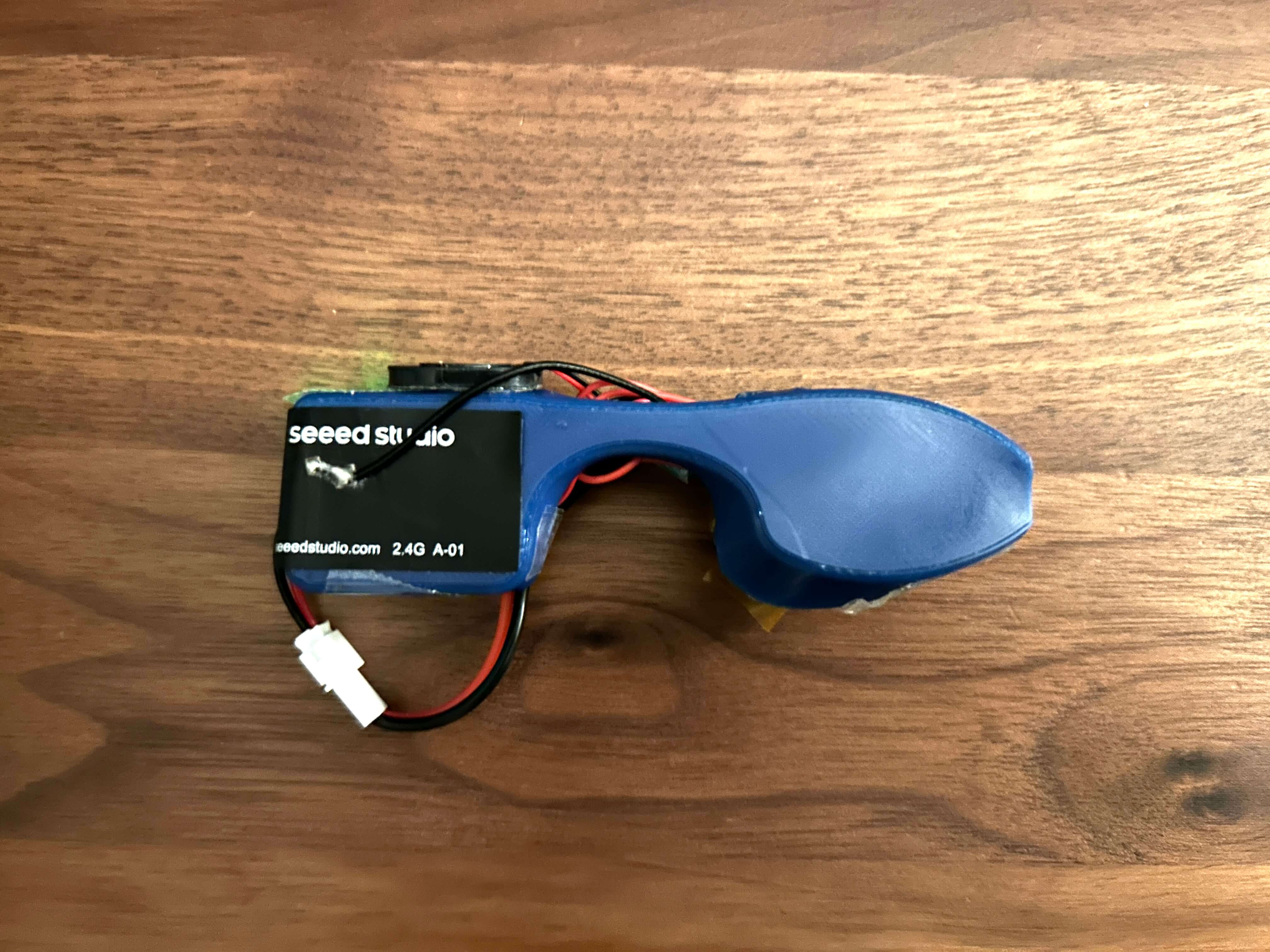}
    \caption{Front}
    \label{fig:pen-front}
  \end{subfigure}\hfill
  \begin{subfigure}{0.48\linewidth}
    \centering
    \includegraphics[
      width=\linewidth,
      trim=0mm 40mm 0mm 75mm,
      clip
    ]{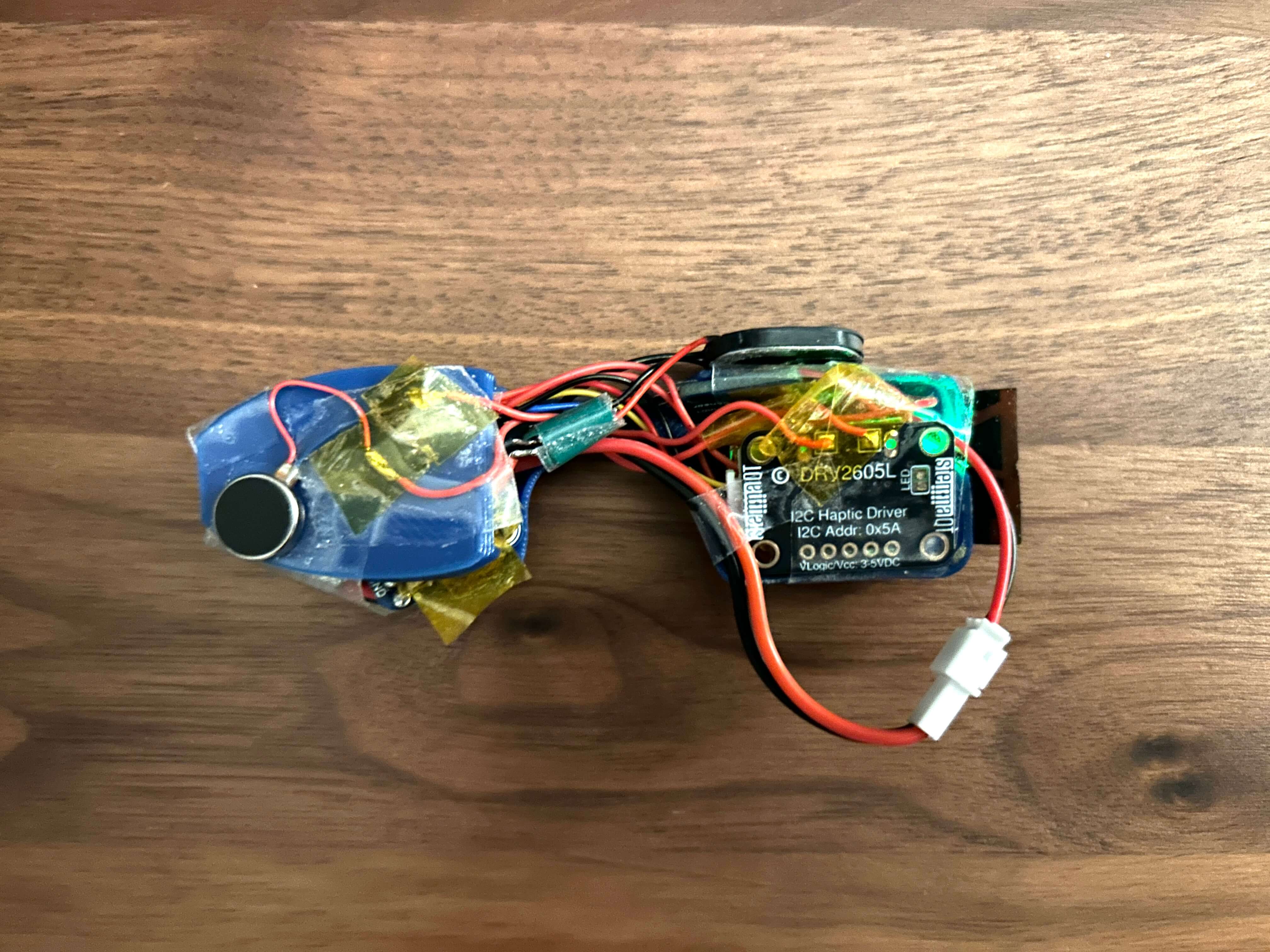}
    \caption{Back}
    \label{fig:pen-back}
  \end{subfigure}
  \caption{Front and back of The Pen prototype.}
  \Description{Two images of The Pen prototype. The left image shows the front view of the pen-shaped ear-worn device. The right image shows the back view, including the haptic motor, speaker, and exposed amplifier.}
  \label{fig:pen-two-up}
\end{figure}

The Pen is a wearable device that resembles a pen or pencil worn behind the user's ear, designed for hands-free interaction during focused work episodes. The device consists of a microphone, camera, speaker, Force Sensitive Resistor (FSR), haptic motor, and UFL antenna. The haptic motor and force sensor enable single press to capture a photo and press-and-hold to initiate a voice query, with haptic feedback communicating system state. Wearing and removing the device marks the beginning and end of an assistance episode. During a session, users interact through voice while remaining engaged in the task (e.g., brainstorming at a whiteboard, washing dishes, automobile maintenance, furniture assembly), with the device capturing visual context and generating task-relevant support such as clarifications or summaries. Additional technical details are described in prior work \cite{tussa2025lessonslearneddevelopingprivacypreserving}.

\section{Study}
\label{sec:study}

We conducted an exploratory study with 6 participants (4 male, 2 female; ages 18--53) across two tasks: (1) reading/studying and (2) whiteboarding/brainstorming. Participants were recruited through local outreach and had no prior experience with the device. Sessions lasted 12--30 minutes. Participants were first introduced to the device and shown how to wear and use it. They were instructed to treat wearing the device as the start of an assistance episode and removal as its end. During each task, participants were encouraged to invoke the device whenever they desired clarification or support. The system provided spoken responses based on captured audio and visual context. Immediately after each session, survey responses and open-ended reflections were collected to understand perceptions of agency, flow, and social comfort.
Because comfort with speaking aloud, perceived recording, and ear-worn devices can vary across demographics and settings, these social acceptability findings should be interpreted as exploratory. Future work should test whether episodic boundaries reveal similar results in more diverse groups and contexts.

\section{Discussion}
\label{sec:discussion}

Our findings show how episodic assistance in a wearable form factor shapes perceptions of agency, trust, flow, and social comfort. Across participants, The Pen was experienced as promising but not without friction, revealing tensions between intentional activation and desires for seamless intelligence. Ending an episode was naturally legible via removal, while starting an episode was often ambiguous and anchored to explicit activation and feedback. We interpret this as evidence that episodic interaction requires layered boundary cues rather than relying on the act of wearing alone.

\subsection{Episodic Boundaries and Agency}

\begin{table}[t]
\centering
\small
\begin{tabular}{p{0.22\linewidth} p{0.7\linewidth}}
\textbf{Boundary} & \textbf{Role} \\
\hline
Physical & Wear/remove establishes start/end of an episode \\
Interactional & Press triggers photo capture; press-and-hold initiates a voice query \\
Perceptual & Haptic response communicates system state \\
\end{tabular}
\caption{Layered boundaries used to stabilize user interpretation of system state.}
\label{tab:boundarylayers}
\end{table}

Participants did not consistently interpret wearing the device as the beginning of an assistance episode. One participant noted that the episode only felt active after pressing the device, suggesting that physical wearing alone was insufficient to signal activation. Participants implicitly sought additional cues such as sounds or haptic feedback to clarify when an episode had begun. These observations indicate that episodic interaction requires explicit feedback to stabilize user understanding of system state. In contrast, removing the device was consistently perceived as a clear end to the interaction: taking it off felt definitive and relieved any sense of ongoing system involvement.

Participants also experienced episodic use as increasing their sense of agency and intentional control over the system. Several preferred using the device only at selected moments rather than continuously, describing it as something that felt “extra” rather than always necessary. Voice interaction was perceived as natural and expressive, enabling participants to articulate questions without shifting attention to a phone and reducing reliance on manual interaction. Non-native English speakers perceived particular value in voice interaction, noting that handheld interfaces often introduce friction when interacting with systems designed primarily in a non-native language.

However, this sense of control came with interaction overhead. Participants reported friction when invoking the device, especially when visual capture required repeated attempts. Technical limitations occasionally disrupted task flow, with sensing failures forcing participants to pause their activity and reducing their ability to remain focused in some cases. As a result, the device was perceived as conceptually supportive of cognitive flow but practically fragile. While physical integration and volitional activation can enhance agency and reduce screen dependence, they also introduce new forms of cognitive and physical effort when sensing and system feedback are unreliable.

% \subsection{Deixis in Episodic Wearables}

\subsection{Legibility, Social Comfort, and the Limits of Episodic Interaction}

Participants expressed low comfort using the device in shared settings, primarily due to concerns about audibility and public perception of wearable cameras. At the same time, participants expressed interest in more proactive capabilities, revealing an unresolved tension between intentional control and seamless intelligence. This suggests that episodic interaction does not eliminate the challenges of wearable sensing, but reframes them as balancing the tradeoffs between the utility of always-on systems and the comfort and control provided by episodic boundaries.

\subsection{Future Work}

Future work should study longer-term use across different settings, including how session boundaries affect personal habits. Additional research is needed on bystander interpretation and social signaling, and perception of alternative boundary-setting rituals (e.g., spatial zones, head motion, wake word activation). Future work should also explore additional signals and gestures that may help stabilize episodic boundaries. For example, pointing, gaze direction \& blink rate via EOG, or subtle hand gestures could function as context-sensitive markers of engagement. Finally, future systems may explore how assistants can better adapt to different types of task episodes while preserving user control and legibility.

Episodic wearable assistance may be particularly meaningful for people with visual impairments. Prior work such as LLM-Glasses demonstrates how AI wearables can support navigation and situational awareness through continuous multimodal feedback \cite{tokmurziyev2026llmglasses}. Our findings suggest an alternative paradigm in which visually impaired users wear a device only when needed to invoke assistance (e.g., during navigation, shopping, etc.), balancing accessibility with autonomy rather than defaulting to always-on support.

\section{Conclusion}
\label{sec:conclusion}

We presented The Pen and an exploratory study of episodic wearable assistance framed around intentional session boundaries. Our findings surface design requirements and tradeoffs that become salient when assistance is intentionally bounded:
(1) start boundaries often require explicit activation and perceptible feedback beyond simply wearing the device;
(2) end boundaries can be naturally and socially interpretable through removal; and
(3) episodic interaction introduces a tradeoff between proactive assistance and user agency. Episodic interaction appears especially promising in contexts where users want short bursts of help without persistent sensing (e.g., shared workspaces, hands-busy tasks), but some users may still value more proactive assistance when reliability and legibility can be preserved. This multi-layered boundary model can help future wearable assistants better align with user preferences for agency, social comfort, and understanding of system state.

\begin{acks}
We thank our study participants for their time and feedback. We also thank our advisor Dr.\ Nirupam Roy for guidance and support throughout this project. This work was partially supported by NSF CAREER Award \#2238433, its associated REU Supplement, and Meta Research Awards.

\end{acks}

\bibliographystyle{ACM-Reference-Format}
\bibliography{references}

\end{document}